\documentclass[twocolumn,showpacs,preprintnumbers,amsmath,amssymb,prb,floatfix]{revtex4}

\usepackage{epsfig,psfrag}
\usepackage{dcolumn}
\usepackage{bm}
\usepackage{graphicx}
\usepackage{color}

\begin{document}

\title{Critical Josephson current through a bistable single-molecule junction}

\author{Andreas Schulz, Alex Zazunov, and Reinhold Egger}

\affiliation{Institut f\"ur Theoretische Physik, 
Heinrich-Heine-Universit\"at, D-40225 D\"usseldorf, Germany}

\date{\today}

\begin{abstract}
We compute the critical Josephson current through a single-molecule
junction. As a model for a molecule with a bistable conformational
degree of freedom, we study an interacting single-level quantum dot
coupled to a two-level system and weakly connected to two superconducting
electrodes. We perform a lowest-order perturbative calculation of
the critical current and show that it can significantly change
due to the two-level system. In particular, the $\pi$-junction behavior,
generally present for strong interactions, can be completely suppressed. 
\end{abstract}

\pacs{74.50.+r, 74.78.Na, 73.63.-b, 85.25.Cp}

\maketitle
\section{Introduction}

The swift progress in molecular electronics achieved during the past
decade has mostly been centered around a detailed understanding of
charge transport through single-molecule junctions,\cite{nitzan,molel}
where quantum effects generally turn out to be important. When two
\textit{superconducting} (instead of normal) electrodes with the same
chemical potential but a phase difference $\varphi$ are attached
to the molecule, the Josephson effect\cite{golubov} implies that
an equilibrium current $I(\varphi)$ flows through the molecular junction.
Over the past decade, experiments have observed gate-tunable Josephson
currents through nanoscale junctions, \cite{helene,morpurgo,reulet,buit1,buit2,doh,kasumov,jorgensen,lieber,leo,CPR-exp1,wernsdorfer,chauvin,christian,nygard,CPR-exp2,landman,hakonen,parisbasel}
including out-of-equilibrium cases, and many different interesting
phenomena have been uncovered. In particular, the current-phase relation
has been measured by employing a superconducting quantum interference
device.\cite{CPR-exp1,wernsdorfer,CPR-exp2} For weakly coupled electrodes,
the current-phase relation is\cite{golubov} \begin{equation}
I(\varphi)=I_{c}\sin(\varphi),\label{cpr}\end{equation}
 with the \textit{critical current} \ $I_{c}$.

The above questions have also been addressed
by many theoretical works. It has been shown that the repulsive
electron-electron (e-e) interaction $U>0$, acting on electrons occupying
the relevant molecular level, can have a major influence on the Josephson
current.\cite{shiba,glazman,spivak,rozhkov,zachar,clerk,matsumoto,vecino,siano,choi,kroha,novotny,meden,florens}
For intermediate-to-strong coupling to the electrodes, an interesting
interplay between the Kondo effect and superconductivity
 takes place.\cite{glazman,clerk,vecino,siano,choi,meden}
In the present paper, we address the opposite limit where, for sufficiently
large $U$, a so-called $\pi$-phase can be realized, with $I_{c}<0$
in Eq.~(\ref{cpr}). In the $\pi$-regime, $\varphi=\pi$ corresponds
to the ground state of the system (or to a minimum of the free energy
for finite temperature), in contrast to the usual $0$-state with
$I_{c}>0$, where $\varphi=0$ in the ground state.\cite{footnew} The sign change
in $I_{c}$ arises due to the blocking of a direct Cooper pair exchange
when $U$ is large. Double occupancy on the molecular level is then
forbidden, and the remaining allowed processes generate the sign change
in $I_{c}$.\cite{shiba,glazman,spivak,zachar,matsumoto,novotny}
The most natural way to explain the $\pi$-junction behavior is by
perturbation theory in the tunnel couplings connecting the molecule
to the electrodes. Experimental observations of the $\pi$-phase
were recently reported for InAs nanowire dots\cite{CPR-exp1} 
and for nanotubes,\cite{wernsdorfer,tomas1} but
a $\pi$-junction is also encountered in 
superconductor-ferromagnet-superconductor
structures.\cite{SFS1,SFS2} Accordingly, theoretical works have also
analyzed spin effects in molecular magnets 
coupled to superconductors.\cite{molmag1,molmag2,molmag3}

The impressive experimental control over supercurrents through molecular
junctions reviewed above implies that modifications of the supercurrent
due to vibrational modes of the molecule play a significant and observable
role.\cite{novotny,bcs-holstein1,bcs-holstein2,bcs-holstein3} We
have recently discussed how a \textit{two-level system} (TLS) coupled
to the dot's charge is affected by the Josephson current carried by 
 Andreev states.\cite{zazuprl}
For instance, two conformational configurations of a molecule may
realize such a TLS degree of freedom. Experimental results for molecular
break junctions with normal leads were interpreted using such models,
\cite{tls-exp1,tls-exp2,tls-exp3,tls-the,tls-the2,fabrizio} but the
TLS can also be created artifically using a Coulomb-blockaded double
dot.\cite{zazuprl} A detailed motivation for our model, where  
the Pauli matrix $\sigma_{z}$ in TLS space couples to the dot's charge,
and its experimental relevance has been given in 
Refs.~\onlinecite{zazuprl,fabrizio}.
While our previous work\cite{zazuprl} studied the Josephson-current-induced
switching of the TLS, we here address a completely different parameter
regime characterized by weak coupling to the electrodes, and focus
on the Josephson current itself. We calculate the critical current
$I_{c}$ in Eq.~(\ref{cpr}) using perturbation theory in these couplings,
allowing for arbitrary e-e interaction strength and TLS parameters.
A similar calculation has been reported recently,\cite{novotny}
but for a harmonic oscillator (phonon mode) instead of the TLS.
Our predictions can be tested experimentally in molecular break junctions
using a superconducting version of existing\cite{tls-exp1,tls-exp2,tls-exp3}
setups.

The remainder of this paper has the following structure. In Sec.~\ref{sec2},
we discuss the model and present the general
perturbative result for the critical current. 
For tunnel matrix element $W_{0}=0$ between the two TLS
states, the result allows for an elementary interpretation,
which we provide in Sec.~\ref{sec3}. The case $W_{0}\ne0$ is then
discussed in Sec.~\ref{sec4}, followed by 
some conclusions in Sec.~\ref{sec5}. Technical details related to
Sec.~\ref{sec3} can be found in an Appendix.
We mostly use units where $e=\hbar=k_{B}=1$.

\section{Model and perturbation theory} \label{sec2}

We study a spin-degenerate molecular dot level with single-particle
energy $\epsilon_{d}$ and on-site Coulomb repulsion $U>0$, coupled
to the TLS and to two standard $s$-wave BCS superconducting banks
(leads). The TLS is characterized by the (bare) energy difference
$E_{0}$ of the two states, and by the tunnel matrix element $W_{0}$.
The model Hamiltonian studied in this paper is motivated by 
Refs.~\onlinecite{tls-exp1,fabrizio} where it was employed to successfully
describe break junction experiments (with normal-state leads). It reads
\begin{equation}
H=H_{0}+H_{{\rm tun}}+H_{{\rm leads}},
\end{equation}
where the coupled dot-plus-TLS part is 
\begin{equation} \label{h0}
H_{0}=-\frac{E_{0}}{2}\sigma_{z}-\frac{W_{0}}{2}\sigma_{x}+
\left(\epsilon_{d}+\frac{\lambda}{2}\sigma_{z}\right)
(n_{\uparrow}+n_{\downarrow})+Un_{\uparrow}n_{\downarrow}
\end{equation}
 with the occupation number $n_{s}=d_{s}^{\dagger}d_{s}^{}$ for 
dot fermion $d_{s}$ with spin $s=\uparrow,\downarrow$. 
Note that the TLS couples with strength $\lambda$ to the dot's
charge. Indeed, assuming some reaction coordinate $X$ describing
molecular conformations, the dipole coupling to the dot 
is $\propto X(n_{\uparrow}+n_{\downarrow})$, just as for electron-phonon
couplings.\cite{bcs-holstein1,bcs-holstein2,tls-exp1,tls-exp2} If
the potential energy $V(X)$ is bistable, the low-energy dynamics
of $X$ can be restricted to the lowest quantum state in each well
and leads to Eq.~\eqref{h0}.
The TLS parameters and the dipole coupling energy $\lambda$ 
can be defined in complete analogy to Refs.~\onlinecite{tls-exp1,fabrizio},
and typical values for $\lambda$ in the meV range are expected,
comparable to typical charging energies $U$. 
Moreover, the electron operators 
$c_{{\bm k}\alpha s}$,
corresponding to spin-$s$ and momentum-${\bm k}$ states in lead $\alpha=L/R$,
are governed by a standard BCS Hamiltonian with complex order parameter
$\Delta_{L/R} e^{\pm i\varphi/2}$ (with $\Delta_{L/R}>0$), 
respectively,
\begin{eqnarray} \label{hleads}
H_{{\rm leads}}&=&\sum_{{\bm k}\alpha s}\epsilon_{{\bm k}\alpha}
c_{{\bm k}\alpha s}^{\dagger}c_{{\bm k}\alpha s}^{} \\
\nonumber &-& \sum_{{\bm k}\alpha}\left(e^{i\alpha\varphi/2}
\Delta_{\alpha}c_{{\bm k}\alpha\uparrow}^{\dagger}
c_{-{\bm k},\alpha\downarrow}^{\dagger}+{\rm h.c.}\right),
\end{eqnarray}
where $\epsilon_{{\bm k}\alpha}$ is the (normal-state) dispersion relation.
Finally, the tunneling Hamiltonian is 
\begin{equation}
H_{{\rm tun}}=\sum_{\alpha s}
\left(H_{T\alpha s}^{(-)}+H_{T\alpha s}^{(+)}\right), \quad
H_{T\alpha s}^{(-)}=\sum_{{\bm k}} t_{{\bm k}\alpha}^{}
c_{{\bm k}\alpha s}^{\dagger}d^{}_{s},
\end{equation}
where $H_{T\alpha s}^{(-)}$ describes tunneling of an electron with spin $s$
from the dot to lead $\alpha$ with tunnel amplitude $t_{{\bm k} \alpha}$, and 
the reverse process is generated by 
$H_{T\alpha s}^{(+)}=H_{T\alpha s}^{(-) \dagger}$.

The Josephson current $I(\varphi)$ at temperature $T=\beta^{-1}$
follows from the equilibrium (imaginary-time) average, 
\begin{equation} \label{jos1}
I=2\ {\rm Im}\left\langle {\cal T}
 e^{-\int_{0}^{\beta}d\tau\ H_{\rm tun}(\tau)}
H_{T\alpha s}^{(-)} \right\rangle ,
\end{equation}
where $\alpha=L/R$ and $s=\uparrow,\downarrow$ can be chosen arbitrarily
by virtue of current conservation and spin-$SU(2)$ invariance, and ${\cal T}$
is the time-ordering operator. Equation
(\ref{jos1}) is then evaluated by lowest-order perturbation theory
in $H_{{\rm tun}}$. The leading contribution is of fourth order in
the tunnel matrix elements and can be evaluated in a similar manner as 
in Ref.~\onlinecite{novotny}. We assume the usual wide-band
approximation for the leads with ${\bm k}$-independent tunnel matrix elements,
and consider temperatures well below both BCS gaps, $T\ll \Delta_{L,R}$.
Putting $\alpha=L$ and $s=\uparrow$, after some algebra,
the Josephson current takes the form (\ref{cpr}) with the critical current  
\begin{equation} \label{Ic}
I_{c}= \frac{2}{\pi^2} \int_{|\Delta_{L}|}^{\infty} \frac{\Gamma_L \Delta_L dE}
{\sqrt{E^2-\Delta_L^2}} \int_{|\Delta_{R}|}^{\infty} 
\frac{\Gamma_R\Delta_R dE'}{\sqrt{E^{\prime 2}- \Delta_R^2}} C(E,E').
\end{equation}
We define the hybridizations $\Gamma_\alpha=\pi \rho_F |t_\alpha|^2$, with 
(normal-state) density of states $\rho_F$ in the leads.
The function $C$ in Eq.~(\ref{Ic}) can be decomposed according to 
\begin{equation}\label{cn}
C(E,E')=\sum_{N=0}^{2}C_{N} (E,E'),
\end{equation}
with contributions $C_{N}$ for fixed dot occupation number 
$N=n_{\uparrow}+n_{\downarrow}=\{0,1,2\}$. 
For given $N$, the two eigenenergies (labeled by $\sigma=\pm$) 
of the dot-plus-TLS Hamiltonian $H_{0}$ in Eq.~(\ref{h0}) are
\begin{equation} \label{h0eig}
E_{N}^{\sigma=\pm}= N\epsilon_{d} + U\delta_{N,2}+ \frac{\sigma}{2}\Phi_{N},
\end{equation}
with the scale
\begin{equation}\label{phin}
\Phi_{N}= \sqrt{(E_{0}- N\lambda)^{2}+W_{0}^{2}}.
\end{equation}
The occupation probability for the state $(N,\sigma)$ is 
\begin{equation}\label{pns}
p_{N}^{\sigma}=\frac{1}{Z} e^{-\beta E_{N}^{\sigma}} (1+\delta_{N,1}),
\end{equation}
where $Z$ ensures normalization, $\sum_{N\sigma} p_N^\sigma=1$.
With the propagator
\begin{equation}
G_{\xi}(E)=\frac{1}{E-\xi},
\end{equation}
we then find the contributions $C_N$ in Eq.~(\ref{cn}),
\begin{widetext}
\begin{eqnarray} \label{C0_gen} 
C_{0}(E,E')&=&\sum_{\sigma_{1}\cdots
\sigma_{4}}\Big[\, p_{0}^{\sigma_{2}}\, 
T_{1010}^{\sigma_{1}\sigma_{2}\sigma_{3}\sigma_{4}}\, 
G_{E_{0}^{\sigma_{2}}-E_{1}^{\sigma_{3}}}(E)\,
G_{E_{0}^{\sigma_{2}}-E_{1}^{\sigma_{1}}}(E')\,
 G_{E_{0}^{\sigma_{2}}-E_{0}^{\sigma_{4}}}(E+E')\\
\nonumber &+& 2p_{0}^{\sigma_{4}}\,
 T_{1210}^{\sigma_{1}\sigma_{2}\sigma_{3}\sigma_{4}}\,
G_{E_{0}^{\sigma_{4}}-E_{1}^{\sigma_{1}}}(E)\, 
G_{E_{0}^{\sigma_{4}}-E_{1}^{\sigma_{3}}}(E')\, 
G_{E_{0}^{\sigma_{4}}-E_{2}^{\sigma_{2}}}(0)\Big ],
\end{eqnarray}
\begin{eqnarray} \label{C1_gen} 
C_{1}(E,E')&=&-\sum_{\sigma_{1}\cdots\sigma_{4}}
\Big[\, T_{1210}^{\sigma_{1}\sigma_{2}\sigma_{3}\sigma_{4}}
\Big(p_{1}^{\sigma_{1}}\, G_{E_{1}^{\sigma_{1}}-E_{0}^{\sigma_{4}}}(E) 
\, G_{E_{1}^{\sigma_{1}}-E_{2}^{\sigma_{2}}}(E)\, 
G_{E_{1}^{\sigma_{1}}-E_{1}^{\sigma_{3}}}(E+E')\\ \nonumber 
&+& p_{1}^{\sigma_{3}}\, G_{E_{1}^{\sigma_{3}}-E_{0}^{\sigma_{4}}}(E')\,
 G_{E_{1}^{\sigma_{3}}-E_{2}^{\sigma_{2}}}(E')\,
 G_{E_{1}^{\sigma_{3}}-E_{1}^{\sigma_{1}}}(E+E')\Big) \\ \nonumber 
&+&\frac{p_{1}^{\sigma_{1}}}{2}\, 
T_{1010}^{\sigma_{1}\sigma_{2}\sigma_{3}\sigma_{4}}\, 
G_{E_{1}^{\sigma_{1}}-E_{0}^{\sigma_{4}}}(E)\,
 G_{E_{1}^{\sigma_{1}}-E_{0}^{\sigma_{2}}}(E')\,
 G_{E_{1}^{\sigma_{1}}-E_{1}^{\sigma_{3}}}(E+E') \\ \nonumber &+&
\frac{p_{1}^{\sigma_{2}}}{2}\, 
T_{2121}^{\sigma_{1}\sigma_{2}\sigma_{3}\sigma_{4}}\, 
G_{E_{1}^{\sigma_{2}}-E_{2}^{\sigma_{3}}}(E)\,
G_{E_{1}^{\sigma_{2}}-E_{2}^{\sigma_{1}}}(E')\, 
G_{E_{1}^{\sigma_{2}}-E_{1}^{\sigma_{4}}}(E+E')\Big],
\end{eqnarray}
\begin{eqnarray} \label{C2_gen}
C_{2}(E,E')&=&\sum_{\sigma_{1}\cdots\sigma_{4}}
\Big[\, p_{2}^{\sigma_{1}}\, 
T_{2121}^{\sigma_{1}\sigma_{2}\sigma_{3}\sigma_{4}}\, 
G_{E_{2}^{\sigma_{1}}-E_{1}^{\sigma_{4}}}(E)\,
G_{E_{2}^{\sigma_{1}}-E_{1}^{\sigma_{2}}}(E')\, 
G_{E_{2}^{\sigma_{1}}-E_{2}^{\sigma_{3}}}(E+E') \\ \nonumber &+&
2p_{2}^{\sigma_{2}}\, T_{1210}^{\sigma_{1}\sigma_{2}\sigma_{3}\sigma_{4}}\,
G_{E_{2}^{\sigma_{2}}-E_{1}^{\sigma_{1}}}(E)\,
 G_{E_{2}^{\sigma_{2}}-E_{1}^{\sigma_{3}}}(E')\,
 G_{E_{2}^{\sigma_{2}}-E_{0}^{\sigma_{4}}}(0)\Big].
\end{eqnarray}
\end{widetext}
Here, we have used the  matrix elements 
\begin{equation}\label{tdef}
T_{N_{1}N_{2}N_{3}N_{4}}^{\sigma_{1}\sigma_{2}\sigma_{3}\sigma_{4}}={\rm Tr}
\left(  A_{N_{1}}^{\sigma_{1}}A_{N_{2}}^{\sigma_{2}}A_{N_{3}}^{\sigma_{3}}
A_{N_{4}}^{\sigma_{4}}\right),
\end{equation}
with the $2\times 2$ matrices (in TLS space)
\[
A_{N}^{\pm}=\frac12 \left(1\mp \frac{(E_{0}-
N \lambda)\sigma_z+W_{0}\sigma_x}{\Phi_{N}}\right).
\]
For $T=0$, it can be shown that 
$C_{0}$ and $C_{2}$ are always positive, while $C_{1}$ 
yields a negative contribution to the critical current.  When $C_1$ outweighs
the two other terms, we arrive at the $\pi$-phase with $I_c<0$.

Below, we consider identical superconductors,
 $\Delta_{L}=\Delta_{R}=\Delta$, and assume $\lambda>0$. 
It is useful to define the reference current scale
\begin{equation}\label{i0}
I_0 = \frac{\Gamma_L\Gamma_R}{\Delta^2}  \frac{2e\Delta}{\pi^2\hbar}. 
\end{equation}
Within lowest-order perturbation theory, the hybridizations $\Gamma_L$ and
 $\Gamma_R$ only enter via Eq.~(\ref{i0}) and can thus be different.
Equation (\ref{Ic}) provides a general but rather complicated expression
for the critical current, even when considering the symmetric case
$\Delta_L=\Delta_R$.  In the next section, we will therefore first
analyze the limiting case $W_0=0$.

\section{No TLS tunneling} \label{sec3}

When there is no tunneling between the two TLS states,
$W_{0}=0$, the Hilbert space of the system can be decomposed
into two orthogonal subspaces ${\cal H}_{+}\oplus{\cal H}_{-}$, with
the fixed conformational state $\sigma=\pm$ in each subspace. 
Equation (\ref{h0eig}) then simplifies to 
\begin{equation} \label{ENsigma}
E_{N}^{\sigma} = \left(\epsilon_{d}+
\frac{\sigma\lambda}{2}\right)N+ U\delta_{N,2}-\frac{\sigma E_{0}}{2}.
\end{equation}
One thus arrives at two decoupled copies of the usual interacting dot problem 
(without TLS), but with a shifted dot level $\epsilon_{\sigma}=
\epsilon_{d}+\sigma\lambda/2$ and the ``zero-point'' energy shift
$-\sigma E_{0}/2$.  As a result, the critical current $I_{c}$ 
in Eq.~(\ref{Ic}) can be written as a weighted sum of the 
partial critical currents $I_{c}(\epsilon_{\sigma})$ 
through an interacting dot level (without TLS) at energy $\epsilon_{\sigma}$,
\begin{equation}\label{ic00}
I_{c}=\sum_{\sigma=\pm} p^{\sigma} I_{c}(\epsilon_{\sigma}),
\end{equation}
where $p^{\sigma}=\sum_{N}p_{N}^{\sigma}$ with Eqs.~(\ref{pns})
and (\ref{ENsigma}) denotes the probability for realizing the 
conformational state $\sigma$.  The current $I_c(\epsilon)$ has already 
been calculated in Ref.~\onlinecite{novotny} 
(in the absence of phonons), and has been reproduced here.  
In order to keep the paper self-contained, we
explicitly specify it in the Appendix.

\begin{figure}
\includegraphics[width=0.40\textwidth]{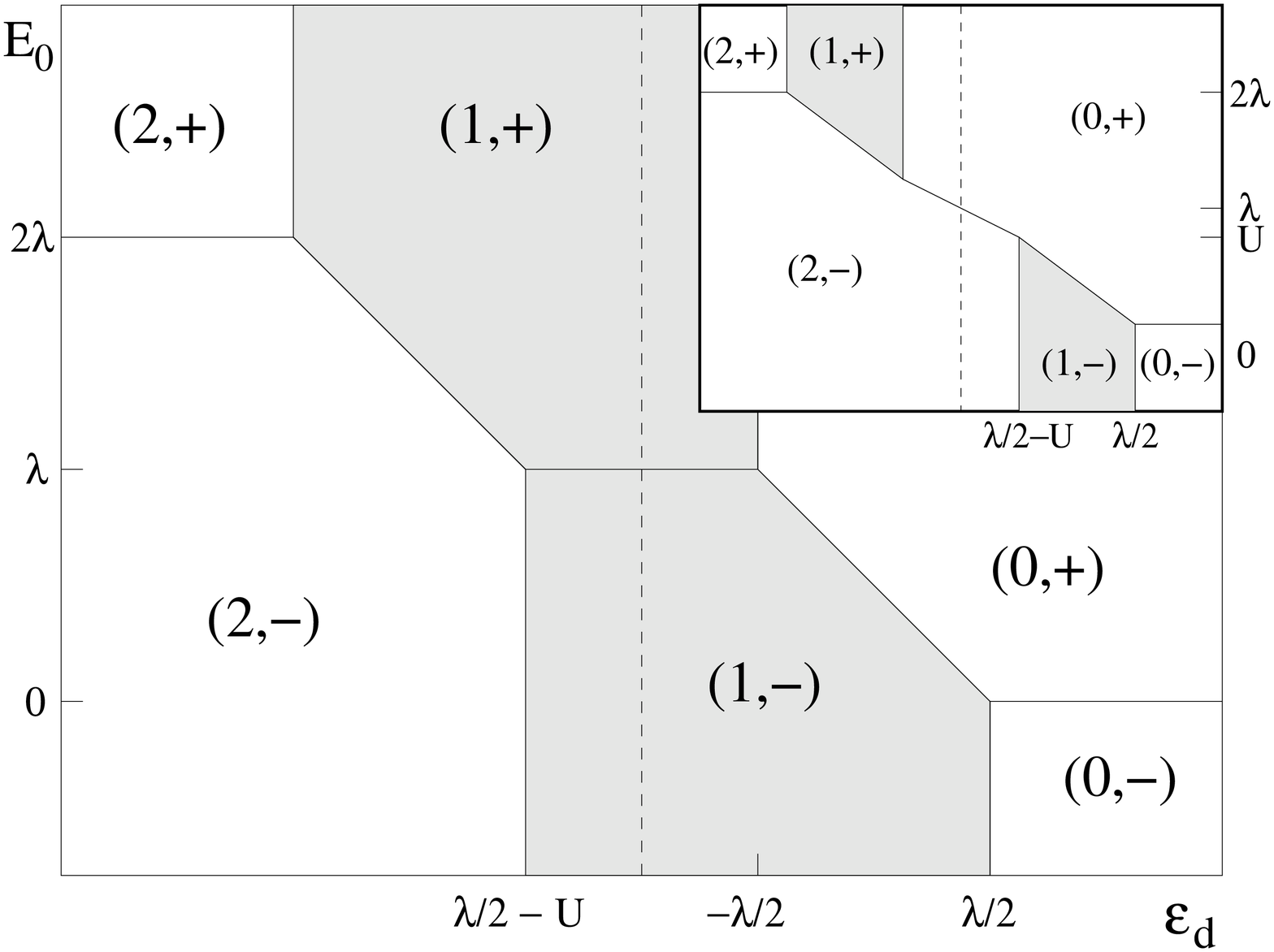}
\caption{\label{fig1} Ground-state phase diagram in the $E_0-\epsilon_d$ 
plane for $W_{0}=0$.  
Different regions $(\bar N,\bar \sigma)$ are labelled according to the 
ground-state dot occupation number $\bar N=0,1,2$ and the conformational 
state $\bar \sigma=\pm$. Dark areas correspond to $\pi$-junction behavior. 
The charge-degeneracy line $\epsilon_d=-U/2$ is indicated as dashed line.
Main panel: $\lambda<U$.  Inset: $\lambda>U$,
where no $\pi$-junction behavior is possible for $U<E_0<2\lambda-U$.  }
\end{figure}

In order to establish the relevant energy scales determining the phase
diagram, we now take the $T=0$ limit. Then the probabilities (\ref{pns})
simplify to $p_{N}^{\sigma}=\delta_{N\bar{N}}\delta_{\sigma\bar{\sigma}}$,
where $E_{\bar{N}}^{\bar{\sigma}}={\rm min}_{(N,\sigma)}\left( 
E_{N}^{\sigma}\right)$ is the ground-state energy of $H_0$
for $W_0=0$.  Depending on the system parameters, 
the ground state then realizes the dot occupation number 
$\bar{N}$ and the TLS state $\bar{\sigma}$. The different 
regions $(\bar N,\bar\sigma)$ in the $E_0-\epsilon_d$ plane 
are shown in the phase diagram in Fig.~\ref{fig1}.  The corresponding critical
current in each of these regions is then simply given by 
$I_{c}=I_{c}(\epsilon_{\bar{\sigma}})$.

\vspace{1cm}
\vspace{1cm}

By analyzing the dependence of the ground-state energy on the system
parameters, one can always (even for $W_0\ne 0$) 
write the function $C(E,E')$ in Eq.~(\ref{Ic}) as 
\begin{eqnarray}\label{eqC}
&& C(E,E')=\Theta(\xi_{-}-\epsilon_{d})\, C_{2}\\ \nonumber
&& + \, \Theta(\epsilon_{d}-\xi_{-}) \, \Theta(\xi_{+}-\epsilon_{d})\, 
C_{1}+\Theta(\epsilon_{d}-\xi_{+})\, C_{0},
\end{eqnarray}
where $\Theta$ is the Heaviside function and the energies
$\xi_{\pm}=\xi_\pm(U,\lambda,E_{0})$ are the boundaries
enclosing the $\pi$-phase region with $\bar N=1$, i.e.,
 $\xi_+$ ($\xi_-$) denotes the boundary between the
$\bar N=0$ and $\bar N=1$ (the $\bar N=1$ and $\bar N=2$) regions,
see Fig.~\ref{fig1}.
Explicit results for $\xi_\pm$ follow from Eq.~(\ref{ENsigma}) for
$W_0=0$.  For $E_0<0$ ($E_0>2\lambda$) and arbitrary $\bar N$, the ground state 
is realized when $\bar \sigma=-$ $(\bar\sigma=+)$, leading to
 $\xi_+=\lambda/2$ ($\xi_+=-\lambda/2$).
In both cases, the other boundary energy follows as $\xi_-=\xi_+-U$.  
In the intermediate cases, with $\xi_0=\frac12(\lambda-U-E_{0}),$
we find for $0<E_{0}<\lambda$, 
\begin{equation}\label{eq1}
\xi_{+}={\rm max}(\lambda/2-E_{0},\xi_0),\quad 
\xi_{-}={\rm min}(\lambda/2-U,\xi_0),
\end{equation}
while for $\lambda<E_0<2\lambda$, we obtain
\begin{equation}\label{eq2}
\xi_{+}={\rm max}(-\lambda/2,\xi_0), \quad
 \xi_{-}={\rm min}(\xi_0,\lambda/2+2\xi_0).
\end{equation}
These results for $\xi_\pm$ are summarized in Fig.~\ref{fig1}.
Remarkably, in the $E_0-\epsilon_d$ plane, the
phase diagram is inversion-symmetric with respect to 
the point $(E_0=\lambda,\epsilon_d=-U/2)$. 
Furthermore, we observe that for many choices of
$E_0$, one can switch the TLS
between the $\bar\sigma=\pm$ states by varying
$\epsilon_d$, see Fig.~\ref{fig1}.

\vspace{1cm}
\begin{figure}
\includegraphics[width=0.45\textwidth]{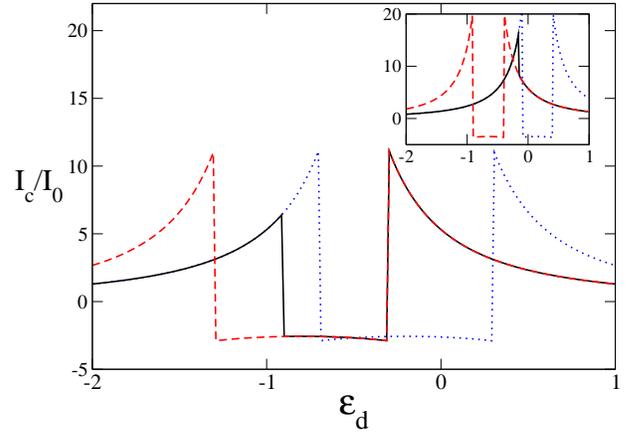}
\caption{\label{fig2} (Color online) Ground-state critical current 
$I_c$ as a function of $\epsilon_d$ for $W_0=0$.  
$I_c$ is given in units of $I_0$, see  Eq.~(\ref{i0}). 
In all figures, the energy scale is set by $\Delta=1$.
Dashed (red), dotted (blue) and solid (black) curves represent the partial 
critical currents $I_{c}(\epsilon_{+})$, $I_{c}(\epsilon_-)$, and 
the realized critical current $I_c$, respectively.
Main panel: $E_{0}=0.8, \lambda=0.6, U=1$, such that
$\xi_+>\xi_-$. This corresponds to the $\pi$-phase region with
$\lambda < E_0 < 2\lambda$ in the main panel of Fig.~\ref{fig1}. 
Inset: $E_{0}=0.6,\lambda=0.8, U=0.5$,
where $\xi_+< \xi_-$ and no $\pi$-junction behavior is possible.
This corresponds to $U<E_0< 2\lambda-U$, see
inset of Fig.~\ref{fig1}. }
\end{figure}
\vspace{1cm}

We now notice that Eq.~(\ref{eqC}) implies the same decomposition for
the critical current  (\ref{Ic}). We can therefore immediately conclude
that the $\pi$-junction regime (where $\bar{N}=1)$ can exist only 
when $\xi_{+}>\xi_{-}$.  This condition is always met away from 
the window $0<E_0<2\lambda$. However, inside that window,
Eqs.~(\ref{eq1}) and (\ref{eq2}) imply that for
sufficiently strong dot-TLS coupling, $\lambda>U$, the $\pi$-phase 
may disappear completely.  Indeed, for $U<E_0< 2\lambda-U$, no
$\pi$-phase is possible for any value of $\epsilon_d$ once $\lambda$ exceeds
$U$.  The resulting ground-state critical current is shown 
as a function of the dot level $\epsilon_d$ for two 
typical parameter sets in Fig.~\ref{fig2}.
The inset shows a case where the $\pi$-phase has been
removed by a strong coupling of the interacting dot to the TLS. 
The above discussion shows that the $\pi$-junction regime is very 
sensitive to the presence of a strongly coupled TLS. 

\section{Finite TLS tunneling} \label{sec4}

Next we address the case of finite TLS tunneling, $W_0 \ne 0$.
Due to the $\sigma_x$ term in $H_0$, the critical current cannot
be written anymore as a weighted sum, see Eq.~(\ref{ic00}),  
and no abrupt switching of the TLS happens when changing 
the system parameters.  Nevertheless, we now show that the 
size and even the existence of the $\pi$-phase region
still sensitively depend on the TLS coupling strength 
(and on the other system parameters).
In particular, the $\pi$-phase can again be 
completely suppressed for strong $\lambda$.

\vspace{1cm}
\begin{figure}
\center\includegraphics[width=0.5\textwidth]{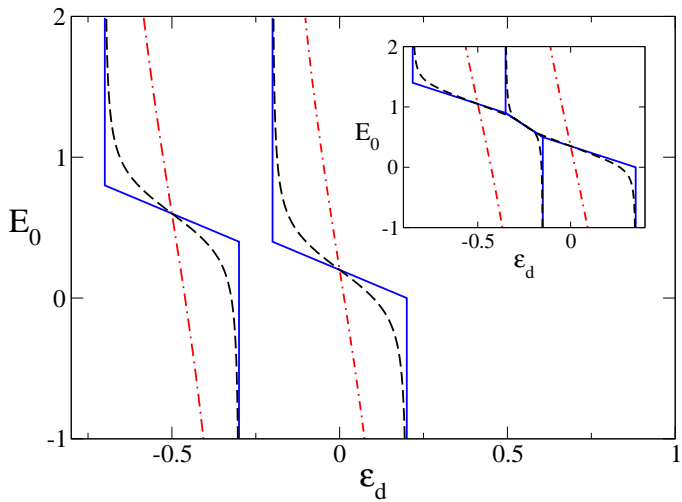}
\caption{\label{fig3} (Color online)
Phase diagram and boundary energies $\xi_{\pm}$ enclosing the $\pi$-phase
for finite $W_{0}$.  Main figure: $\lambda=0.4$ and $U=0.5$, where
a $\pi$-phase is present; $W_0=0, 0.2$ and $5$, 
for solid (blue), dashed (black) and 
dash-dotted (red) curves, respectively. 
Inset: $\lambda=0.7$ and  $U=0.5$, where the $\pi$-phase vanishes;
$W_0=0, 0.3$ and $3$, for solid (blue), dashed (black)
and dash-dotted (red) curves, respectively.}
\end{figure}
\vspace{1cm}

For finite $W_{0}$, the ground-state critical current is obtained
from Eq.~(\ref{eqC}), where
the $C_{N}$ are given by Eqs.~(\ref{C0_gen})--(\ref{C2_gen})
and the $\pi$-phase border energies $\xi_{\pm}$ are replaced by 
\begin{equation} \label{eqB}
\xi_{+}= \frac{1}{2}\left(\Phi_{1}-\Phi_{0}\right),\quad
\xi_{-}= \frac{1}{2}\left(\Phi_{2}-\Phi_{1}-2U\right).
\end{equation}
The $\Phi_N$ are  defined in Eq.~(\ref{phin}).
Compared to the $W_0=0$ case in Fig.~\ref{fig1},
the phase diagram boundaries now have a
smooth (smeared) shape due to the TLS tunneling. Nevertheless,
the critical current changes sign abruptly when the system parameters are 
tuned across such a boundary. 
The energies (\ref{eqB}) are shown in Fig.~\ref{fig3} for various 
values of $W_0$ in the $E_0-\epsilon_d$ plane.  
In between the $\xi_+$ and $\xi_-$ curves, the $\pi$-phase
is realized.  From the inset of Fig.~\ref{fig3}, we indeed confirm that
the $\pi$-phase can again be absent within a suitable
parameter window.  Just as for $W_{0}=0$, the
 $\pi$-phase vanishes for $\xi_{+}<\xi_{-}$,
and the transition between left and right $0$-phase 
occurs at $\bar\xi = (\xi_{+}+\xi_{-})/2$. 
For $|E_{0}|\gg {\rm max}( \lambda, W_0)$,  we effectively recover
the phase diagram for $W_0=0$, since the TLS predominantly occupies
a fixed conformational state.

\vspace{1cm}
\begin{figure}[!h]
\includegraphics[width=0.45\textwidth]{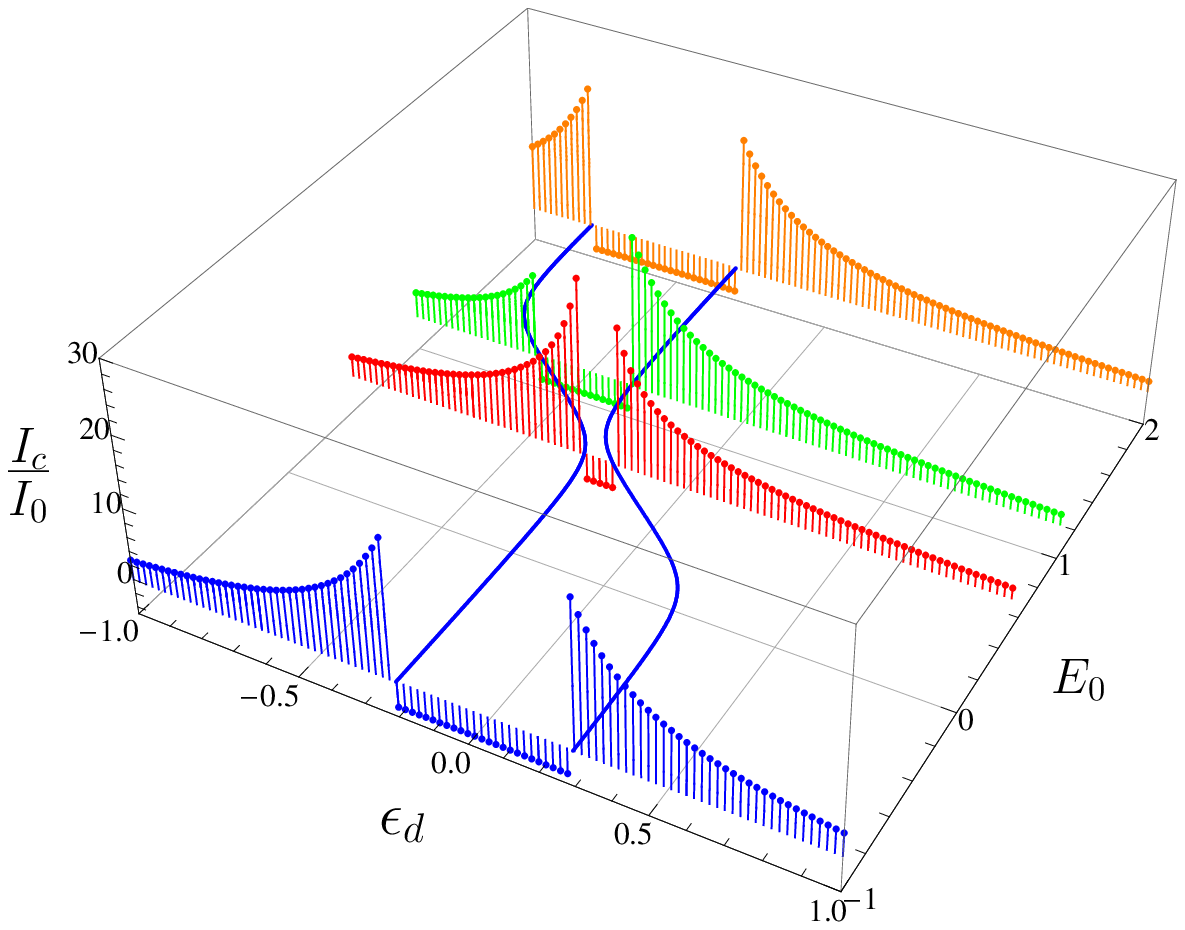}
\includegraphics[width=0.45\textwidth]{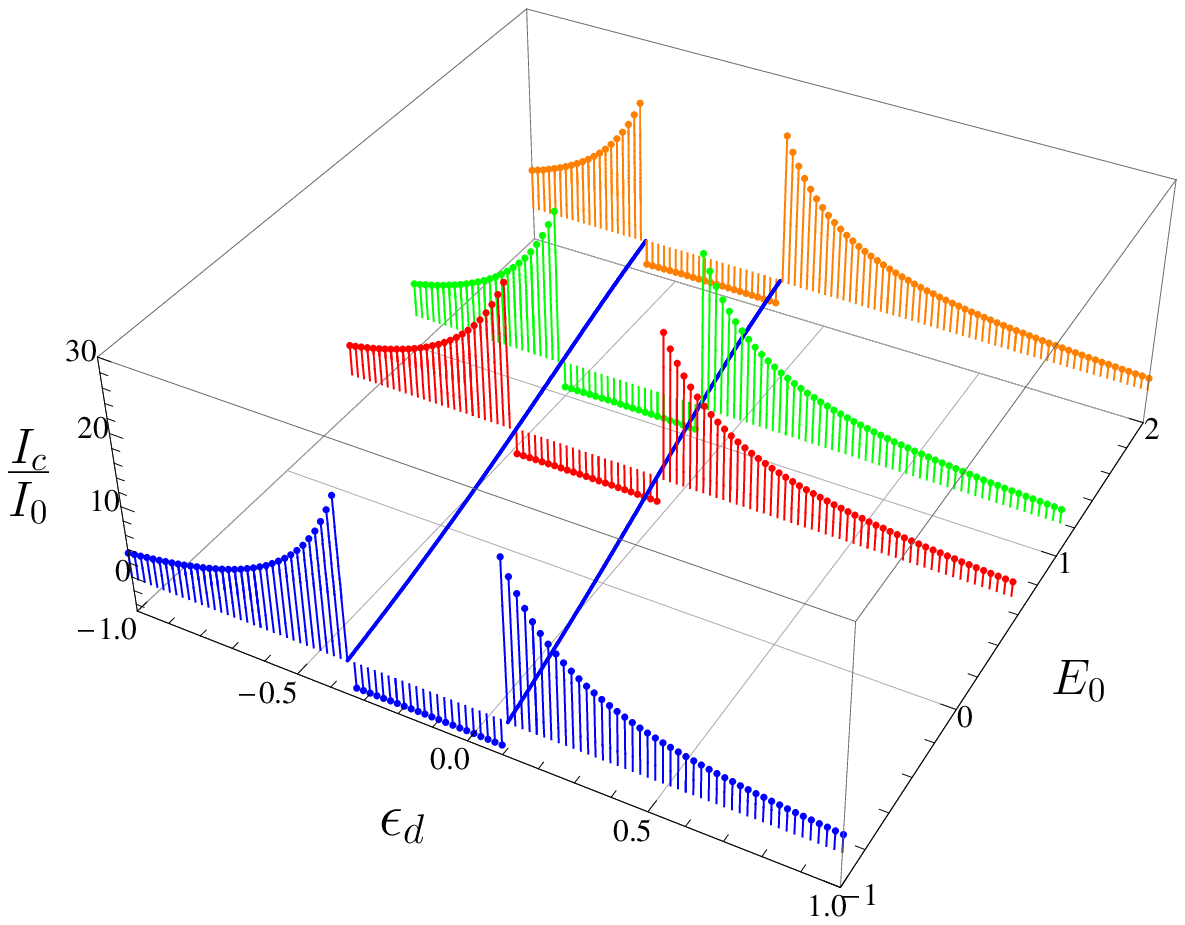}
\caption{\label{fig4} (Color online) Listline plots of 
the $W_0\ne 0$ ground-state critical current $I_c$ (in units of $I_0$)
in the $E_0-\epsilon_d$ plane, with $\lambda=0.6$ and $U=0.5$. 
The boundaries $\xi_\pm$ enclosing the $\pi$-phase, see also Fig.~\ref{fig3},
are indicated as solid (blue) curves.
Top panel: Small tunnel amplitude, $W_0=0.2$. 
Bottom panel: Large tunnel amplitude, $W_{0}=3$.}
\end{figure}
\vspace{1cm}

The corresponding critical current $I_c$ is shown
in Fig.~\ref{fig4} for both a small and a very large TLS tunnel matrix
element $W_0$. 
In the limit of large $W_{0}\gg{\rm max}(\lambda,|E_{0}|)$, 
see lower panel in Fig.~\ref{fig4}, the dot and the TLS 
are effectively decoupled, since $\langle\sigma_z\rangle \simeq 0$ and 
$\langle\sigma_{x}\rangle\simeq {\rm sgn}( W_{0})$.
While this limit is unrealistic for molecular junctions, it may
be realized in a side-coupled double-dot system.\cite{zazuprl}
Finally we note that, unlike for $W_0=0$, 
the perturbative result for the critical current 
\textit{diverges} at the point where the $\pi$-phase vanishes, i.e.,
for $\epsilon_d=\bar \xi$.  This divergence is an artefact of perturbation
theory and is caused by the appearance of the 
factor $G_{E_{0}^{-}-E_{2}^{-}}(0)=
(\epsilon_d-\bar\xi)^{-1}$ in Eqs.~(\ref{C0_gen}) 
and (\ref{C2_gen}). 

\section{Conclusions} \label{sec5}

In this paper, we have presented a perturbative calculation of the 
critical Josephson current, $I_c$, through an interacting single-level 
molecular junction side-coupled to a two-level system (TLS).
Such a TLS is a simple model for a bistable conformational degree
of freedom, and has previously been introduced in the 
literature.\cite{zazuprl,tls-exp1,fabrizio}
Our perturbative calculation assumes very
weak coupling to attached superconducting reservoirs. 
The ground-state critical current can then be computed exactly for
otherwise arbitrary parameters. Our main finding is that the 
$\pi$-phase with $I_c<0$ is quite sensitive to the presence of 
the TLS.  In particular, for strong coupling $\lambda$ of the molecular level
to the TLS as compared to the Coulomb energy $U$ on the level, 
the $\pi$-phase can disappear altogether.   

\acknowledgments

We thank T. Novotn{\' y} for discussions.
This work was supported by the SFB TR 12 of the DFG and by the EU
networks INSTANS and HYSWITCH.

\appendix
\section{Partial critical currents}

In this Appendix, we provide the partial critical current
$I_c(\epsilon_\sigma)$ which appears in the 
calculation for $W_0=0$, see Sec.~\ref{sec3}. 
In the absence of TLS tunneling, the matrix elements (\ref{tdef})
simplify to
\[
 T_{N_{1}N_{2}N_{3}N_{4}}^{\sigma_{1}\sigma_{2}\sigma_{3}\sigma_{4}}=
\prod_{i=1}^4 \delta_{\tilde\sigma_i,1} +
\prod_{i=1}^4 \delta_{\tilde\sigma_i,-1},
\]
where $\tilde\sigma_i=\sigma_i {\rm sgn}(N_i\lambda-E_0)$.
We now rename $\tilde\sigma\to \sigma$ to denote the conformational
state (eigenstate of $\sigma_z$). 

The partial current $I_{c}(\epsilon_{\sigma})$ corresponding to fixed
conformational state $\sigma=\pm$ is then given by 
\[
\frac{I_{c}(\epsilon_{\sigma})}{I_0}= \Delta^3 \sum_N \int_{\Delta}^{\infty}
\frac{dEdE' C_{N}^{\sigma}(E,E')}
{\sqrt{(E^2-\Delta^2)(E^{\prime 2}-\Delta^2)}},
\]
where 
\begin{eqnarray*}
C_{N}^{\sigma}(E,E')&=& \tilde{p}_{N}^{\sigma}c_{N}^{\sigma}(E,E'),
\\
\tilde{p}_{N}^{\sigma}&=& \frac{1}{Z_\sigma} 
 e^{-\beta E_{N}^{\sigma}} \, (1+\delta_{N,1}),
\end{eqnarray*}
with $Z_{\sigma}$ such that $\sum_N \tilde p_N^\sigma=1$.
Moreover, the $c^\sigma_N$ are given by
\[
c_{0}^{\sigma}(E,E')=\frac{1}{(E+\epsilon_{\sigma})
(E'+\epsilon_{\sigma})}\left[\frac{1}{E+E'}+\frac{2}
{2\epsilon_{\sigma}+U}\right],
\]
\begin{eqnarray*}
&& c_{1}^{\sigma}(E,E')= -\frac{1}{E+E'}\Big [ 
\frac{1}{(E-\epsilon_{\sigma})(E+\epsilon_{\sigma}+U)} \\
&& +\,\frac{1}{(E'-\epsilon_{\sigma})
(E'+\epsilon_{\sigma}+U)}+\frac{1/2}
{(E-\epsilon_{\sigma})(E'-\epsilon_{\sigma})}\\
&& +\,\frac{1/2}{(E+\epsilon_{\sigma}+U)
(E'+\epsilon_{\sigma}+U)}\Big ],
\end{eqnarray*}
\begin{eqnarray*}
c_{2}^{\sigma}(E,E')&=&\frac{1}{(E-\epsilon_{\sigma}-U)
(E'-\epsilon_{\sigma}-U)} \\
&\times&\left[\frac{1}{E+E'}-
\frac{2}{2\epsilon_{\sigma}+U}\right].
\end{eqnarray*}


\begin{thebibliography}{10}
\bibitem{footnew}
For finite $\Gamma/\Delta$, one also has intermediate $0'$ and $\pi'$
phases.\cite{rozhkov} These phases disappear, however, in the limit
$\Gamma/\Delta\ll 1$ considered in this work.
\end{thebibliography}
\end{document}